\documentclass[runningheads,fleqn]{cl2emult}

\usepackage{graphicx} 
\usepackage{subeqnar} 

\begin{document}
\title*{Dynamic Phase Transition and Hysteresis in Kinetic Ising Models}
%
%
%
%
\titlerunning{~}
%
\author{
P.~A.~Rikvold\inst{1,2}
\and 
G.~Korniss\inst{2}
\and 
C.~J.~White\inst{1,2}
\and 
M.~A.~Novotny\inst{2}
\and 
S.~W.~Sides\inst{3}
}
\authorrunning{~}
%
%
\institute{
Department of Physics and 
Center for Materials Research and Technology, 
Florida State University, Tallahassee, FL 32306-4350, USA
\and 
Supercomputer Computations Research Institute, 
Florida State University, Tallahassee, FL 32306-4130, USA
\and 
Integrated Materials Research Laboratory, Sandia National Laboratory, 
Albuquerque, NM 87123, USA
}

\maketitle              

\begin{abstract}
We briefly introduce hysteresis
in spatially extended systems and the 
dynamic phase transition
observed as the frequency of the oscillating field 
increases beyond a critical value. 
Hysteresis and the decay of metastable phases
are closely related  
phenomena, and a dynamic phase transition can occur only for field amplitudes, 
temperatures, and system sizes at which the metastable 
phase decays through nucleation and growth of {\it many\/} droplets.
We present preliminary results from extensive 
Monte Carlo simulations
of a two-dimensional kinetic Ising model in a square-wave 
oscillating field and estimate critical exponents by finite-size scaling
techniques adapted from equilibrium critical phenomena. 
The estimates are consistent with the universality class 
of the two-dimensional 
equilibrium Ising model and inconsistent with two-dimensional random 
percolation.
However, we are not aware of any theoretical arguments indicating 
why this should be so. Thus, the question of the universality 
class of this nonequilibrium critical phenomenon remains open. 
\end{abstract}

\section{Introduction} 
\label{sec:intro} 
\begin{figure}
\centerline{
\hspace*{-1.7truecm}
\rotatebox{270}{
\includegraphics[width=0.5\textwidth]{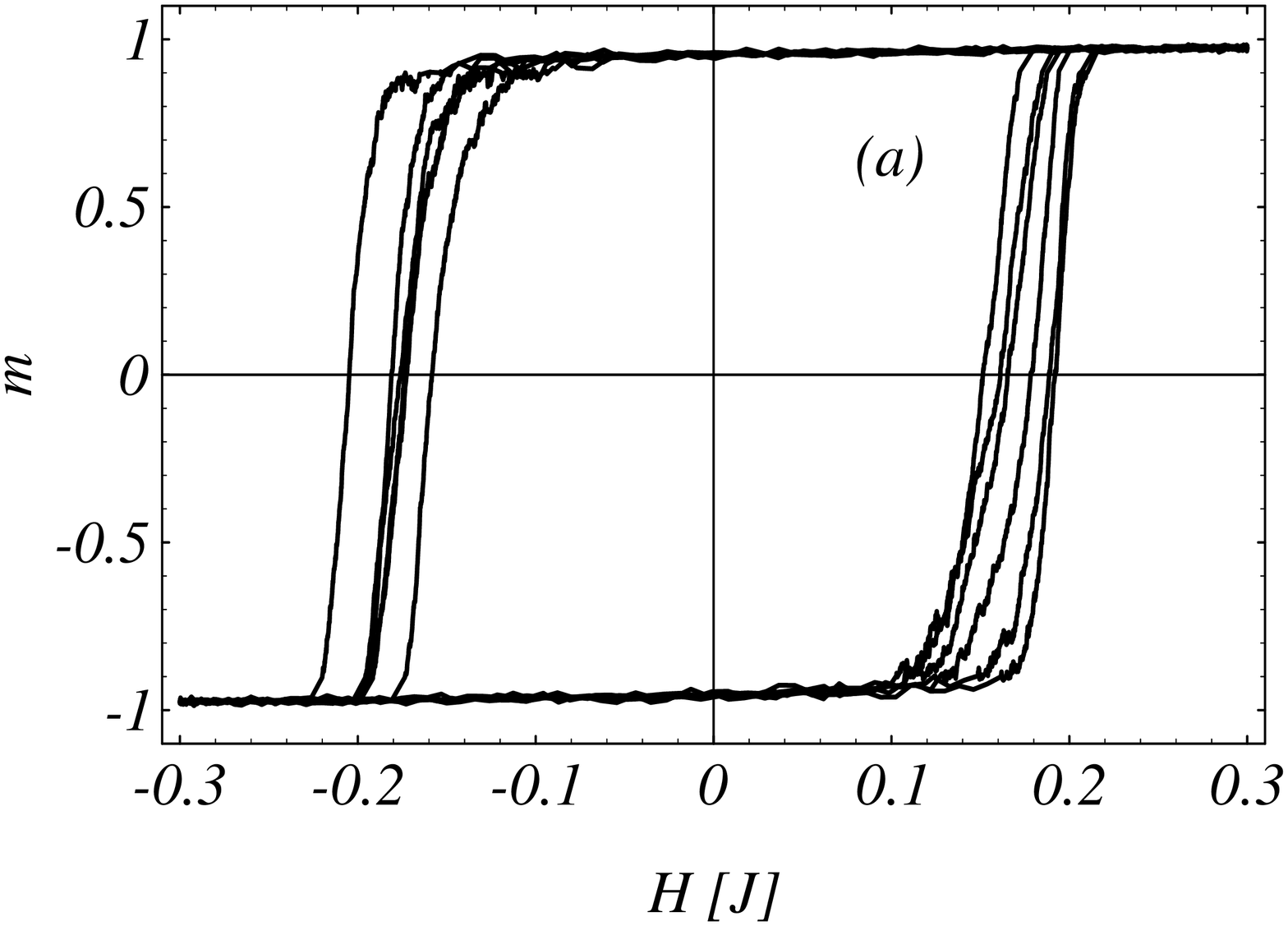}
}
\hspace*{ 1.4truecm}
\rotatebox{270}{
\includegraphics[width=0.5\textwidth]{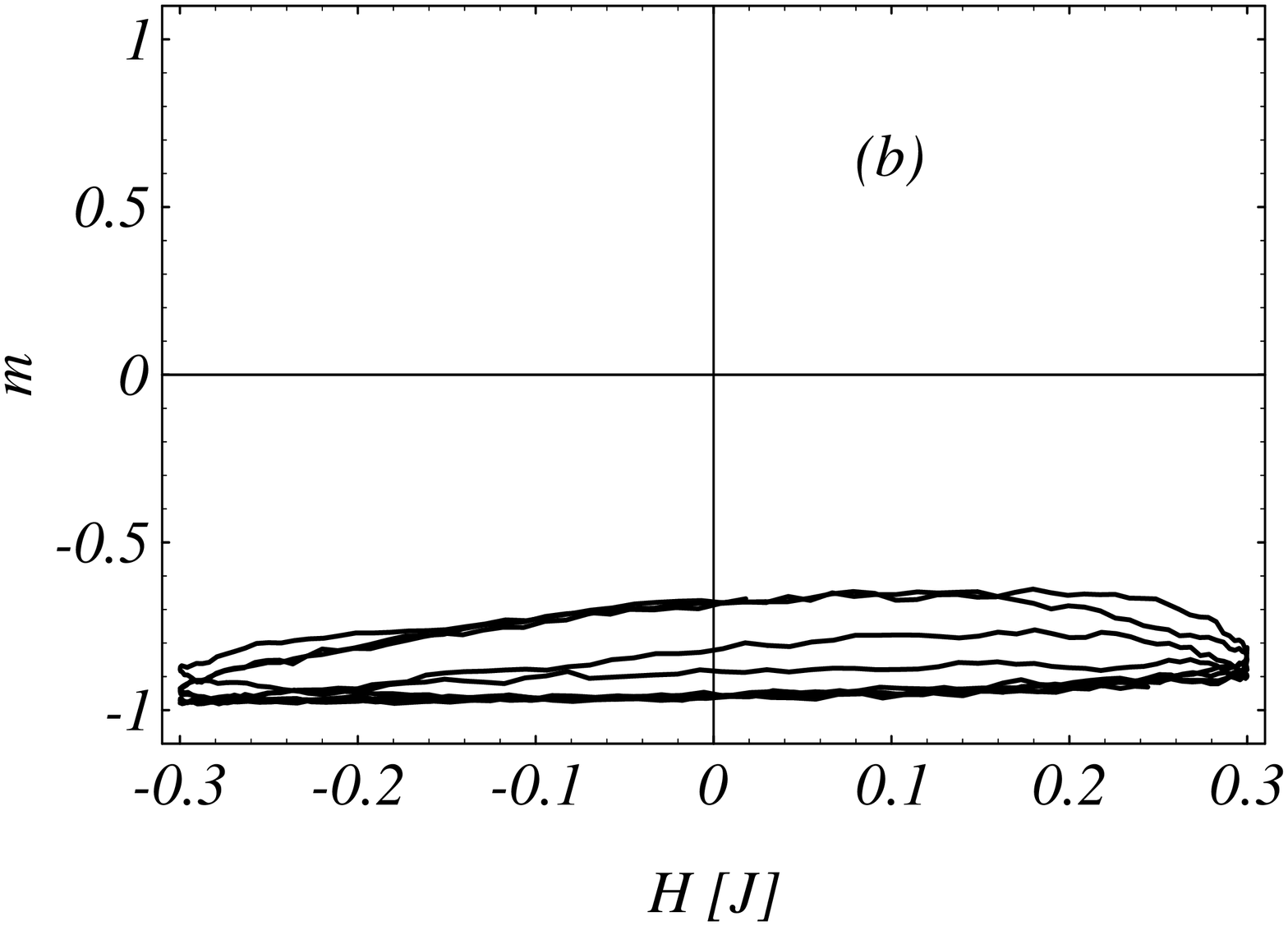}
}
}
\vspace*{-2.0truecm}
\caption[]
{
Series of hysteresis loops obtained from a Monte 
Carlo (MC) simulation of a two-dimensional kinetic Ising ferromagnet at 
$T \! = \! 0.8 \ T_{\mathrm c}$ in a sinusoidally varying field. 
The dimensionless magnetization $m$ is shown 
vs.\ the field $H$ in units of the nearest-neighbor interaction energy $J$. 
(a):
A series of regular symmetric hysteresis loops, typical of a system driven 
at a moderately low frequency. 
(b):
A series of asymmetric hysteresis loops, typical of a system 
driven at a high frequency}
\label{fig:MHloops}
\end{figure}

Hysteresis often occurs when a bistable or multistable 
system is driven by an oscillating 
force which varies too fast for the system to respond without a phase lag. 
In fact, the word was coined by Ewing from the Greek {\it husterein\/} 
\mbox{($\stackrel{\scriptscriptstyle c}{\upsilon} \! \! \sigma \tau
\epsilon \rho \acute{\epsilon} \omega$)}
which means ``to be behind'' \cite{EWIN1881}.
Examples include ferromagnets and ferroelectrics in AC fields, 
electrochemical cyclic-voltammetry
experiments, and nonlinear elastic media
under oscillating stress, just to mention a few. 
The most familiar representation 
of hysteresis is probably the (usually) closed curve 
obtained by plotting the system response versus the applied force. 
Examples of such {\it hysteresis loops\/}
are shown in Fig.~\ref{fig:MHloops}(a). This figure shows data 
from a Monte Carlo (MC) simulation of a two-dimensional kinetic Ising 
ferromagnet 
below its critical temperature, which is driven by a sinusoidally oscillating 
field. However, the shape of the loops shown is quite general. 
For simplicity and concreteness, in this paper we use magnetic language,
designating the oscillating force the ``field'' and the system response the 
``magnetization''.
These quantities can easily be re-interpreted when one discusses other
types of systems, such as those mentioned above.

Hysteresis was first systematically investigated 
in the late 19th century by engineers and physicists 
primarily concerned with the development of electric motors and transformers 
\cite{EWIN1881,WARB1881,EWIN1882,STEI1892,KLIN92}. 
The area of the hysteresis loop is proportional to the 
magnetic energy loss during one field period, 
as was first pointed out by Warburg \cite{WARB1881}. 
Its dependences on the frequency and 
amplitude of the applied field have therefore 
been studied intensively ever since, but open questions still remain. 
In particular, the question of whether the low-frequency behavior 
for ultrathin films of highly anisotropic materials is asymptotically 
a power law or logarithmic is still under investigation, both experimentally 
\cite{HE93,JIAN95,JIAN96,SUEN97,SUEN99}
and theoretically 
\cite{JUNG90,RAO90B,THOM93,LUSE94,BEAL94,SIDE98B,SIDE99}. 

A different aspect of hysteresis in bistable systems, 
which is the main topic of the present 
paper, occurs at higher driving frequencies. 
When the frequency becomes sufficiently high, the symmetry of the 
hysteresis loop, which is apparent in Fig.~\ref{fig:MHloops}(a), is broken. 
Instead of oscillating between its two stable values with a phase lag relative 
to the field, the magnetization oscillates around one or the other of its 
zero-field stable values. 
A series of such asymmetric loops is shown in Fig.~\ref{fig:MHloops}(b). 
This symmetry breaking
has become a topic of vigorous research during the last 
two decades. It was first reported by Tom{\'e} and de~Oliveira 
\cite{TOME90}, who observed it 
during numerical solutions of a mean-field equation of motion 
for a ferromagnet in an oscillating field. 

Subsequently, symmetry breaking  
has been observed in numerous MC simulations of kinetic Ising systems, 
\cite{SIDE99,LO90,ACHA95,ACHA97C,ACHA97D,ACHA98,SIDE98}, 
as well as in further mean-field studies 
\cite{ACHA95,ACHA97D,ACHA98,BUEN98}. 
It may also have been experimentally observed in ultrathin films of 
Co on Cu(001) \cite{JIAN95,JIAN96}. 
Reviews of the field as it stood in 1994 and 
as it stands today can be found 
in \cite{ACHA94} and~\cite{CHAK99}, respectively. 
There now appears to be a consensus that the symmetry breaking 
corresponds to a genuine second-order, 
nonequilibrium phase transition. 
Associated with the transition is a divergent time scale 
(critical slowing-down \cite{ACHA97D} and, 
for spatially extended systems, a divergent 
correlation length \cite{SIDE99,SIDE98}. 
Although estimates of the critical exponents have recently become 
available from finite-size 
scaling analyses of MC data for a two-dimensional Ising system 
in a sinusoidally oscillating field \cite{SIDE99,SIDE98}, their accuracy is 
not yet sufficient to decide whether the dynamic phase transition in this 
system belongs to a previously known universality class or represents a 
new one. Neither are we aware of theoretical arguments that can resolve the 
issue or help to identify the correct ``critical clusters''. 

The main purpose of this paper is to present preliminary 
results for two-dimensional kinetic Ising systems 
in square-wave oscillating fields. These may help point the way towards 
answering some of the remaining questions concerning this intriguing 
nonequilibrium critical phenomenon. The use of a square-wave field both tests
the universality of the dynamic phase transition and significantly
increases the computational speed.
The structure of the remainder of the paper is as follows. 
The kinetic model system and the relevant quantities, 
including the dynamic order parameter, are defined in Sect.~\ref{sec:model}. 
A brief primer on metastable decay in spatially extended systems 
is given in Sect.~\ref{sec:KJMA}. 
Numerical results are presented in Sect.~\ref{sec:res}, and a discussion and 
some suggestions for further research are given in Sect.~\ref{sec:disc}. 

\section{Model and Relevant Quantities} 
\label{sec:model} 
Most numerical simulations of the dynamic phase transition 
(hereafter abbreviated DPT) in spatially extended systems have been 
performed on nearest-neighbor kinetic Ising ferromagnets on hypercubic 
lattices with periodic boundary conditions. These models are defined by 
the Hamiltonian 
\begin{equation}
\label{eq:Hamil}
{\cal H } = -J \sum_{ {\langle ij \rangle}} {s_{i}s_{j}}
               - H(t) \sum_{i} {s_{i}} \;,
\end{equation}
where $s_{i} \! = \! \pm 1$ is the state of the $i$th spin, 
$\sum_{ {\langle ij \rangle} }$ runs over all nearest-neighbor pairs, 
$J > 0$ is the ferromagnetic interaction, 
$\sum_{i}$ runs over all $L^{d}$ lattice sites, and $H(t)$ is an oscillating, 
spatially uniform applied field. The magnetization per site is
\begin{equation}
\label{eq:m(t)} 
{m(t)} = {L^{-d}} \sum_{i=1}^{L^d} {s_{i}(t)} \;.
\end{equation}
The temperature $T$ is fixed below its zero-field critical value 
$T_{\mathrm c}$, so that the magnetization for $H\!=\!0$ has two degenerate 
spontaneous equilibrium values, $\pm m_{\mathrm{sp}}(T)$. For nonzero 
fields the equilibrium magnetization has the same sign as 
$H$, while for not too strong $H$ 
(see Sect.~\ref{sec:KJMA} for quantitative statements) 
the opposite magnetization direction becomes {\it metastable\/} 
and decays slowly towards equilibrium with time. 

The dynamic used here, as well as in \cite{SIDE99} and \cite{SIDE98}, 
is the Glauber single-spin-flip MC algorithm 
with updates at randomly chosen sites \cite{BIND92B}. 
The time unit is one MC step per spin (MCSS). Each attempted spin flip 
from ${s_{i}}$ to ${-s_{i}}$ is accepted with probability 
\begin{equation}
\label{eq:Glauber}
W(s_{i} \rightarrow -s_{i}) =
\frac{ \exp(- \beta \Delta E_{i})}{1 + \exp(- \beta \Delta E_{i})} \; .
\end{equation}
Here $\Delta E_{i}$ is the energy change 
that would result from accepting the spin flip, 
and $\beta \! = \! 1/k_{\rm B}T$ where $k_{\rm B}$ is Boltzmann's constant. 
For the largest system ($L \!=\! 512$ square lattice) we employed a massively
parallel implementation of this algorithm \cite{LUBAA,LUBAB,KBNR,KNR}. 
Other dynamics that have been used in MC studies of the DPT are 
Glauber or Metropolis \cite{BIND92B} 
with updates at sequentially selected sites 
\cite{ACHA95,ACHA97C,ACHA97D,ACHA98,ACHA94,CHAK99}. Although the choice of 
update scheme can lead to subtle differences in the dynamics 
\cite{RIKV94A} and we prefer random site selection as the more physical 
scheme, we do not believe it affects universal aspects of the DPT. 

The dynamic order parameter is the 
period-averaged magnetization \cite{TOME90}, 
\begin{equation}
\label{eq:Qeq}
Q = \frac{1}{2 t_{1/2}} \oint m(t) \ {\mathrm d}t \;,
\end{equation}
where $t_{1/2}$ is the half-period of the oscillating field, and 
the beginning of the period is chosen at a time when $H(t)$ 
changes sign. Analogously we also define the local order parameter
\begin{equation}
\label{eq:Qlocaleq}
Q_{i} = \frac{1}{2 t_{1/2}} \oint s_{i}(t) \ {\mathrm d}t \;,
\end{equation}
which is the period-averaged spin at site $i$. For 
slowly varying fields the probability distribution of $Q$ 
is sharply peaked at zero \cite{SIDE99}. 
We shall refer to this as the {\it dynamically disordered phase\/}. For 
rapidly oscillating fields the $Q$ distribution 
becomes bimodal with two sharp peaks near $\pm m_{\mathrm{sp}}(T)$, 
corresponding to the broken symmetry of the hysteresis loops \cite{SIDE99}. 
We shall refer to this as the {\it dynamically ordered phase\/}.
Near the DPT we use finite-size scaling analysis of 
MC data to estimate the critical exponents that characterize the transition. 

Previous studies of the DPT have used an applied field
which varies sinusoidally in time. While sinusoidal or 
linear saw-tooth fields are the most common in experiments and are 
necessary to obtain a vanishing loop area in the low-frequency 
limit \cite{SIDE98B,SIDE99}, 
the wave form of the field should not affect universal 
aspects of the DPT. This should be so because the transition essentially 
depends on the competition between two time scales: the half-period $t_{1/2}$
of the applied field, and the average time it takes the system to 
leave the metastable region near one of its two degenerate 
zero-field equilibria when a field of magnitude $H_0$ and sign 
opposite to the magnetization is applied. 
This {\it metastable lifetime\/}, $\langle \tau(H_0,T) \rangle$, 
is usually estimated as the average 
first-passage time to zero magnetization. 
In the present paper we use a {\it square-wave\/} field of amplitude $H_0$.
This has significant computational advantages over the sinusoidal 
field variation since we can use two look-up tables to determine the 
acceptance probabilities: one for $H \!=\! + H_0$ and one for $H \!=\! - H_0$.

\begin{figure}
\includegraphics[width=.48\textwidth]{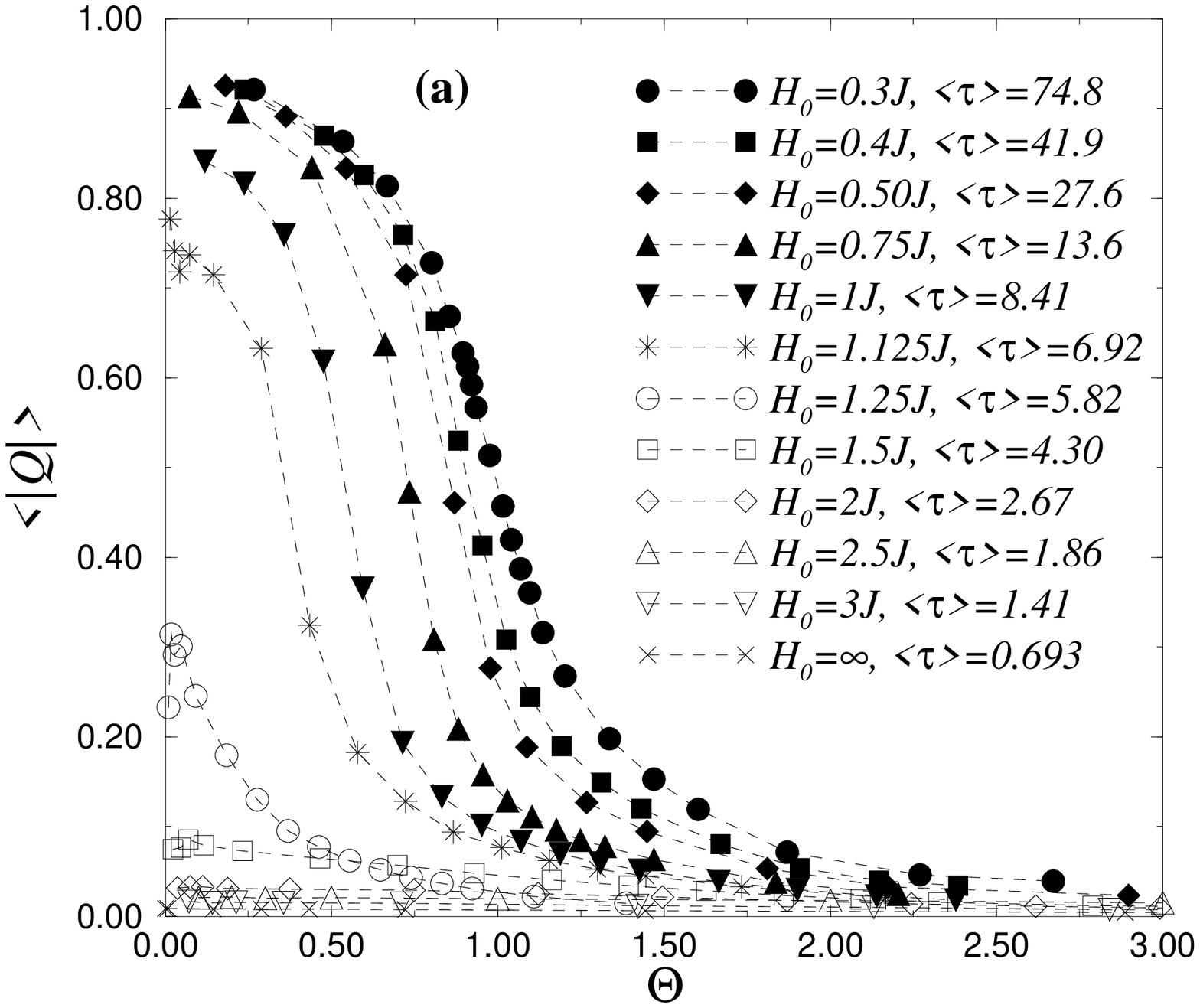} \hfill
\includegraphics[width=.48\textwidth]{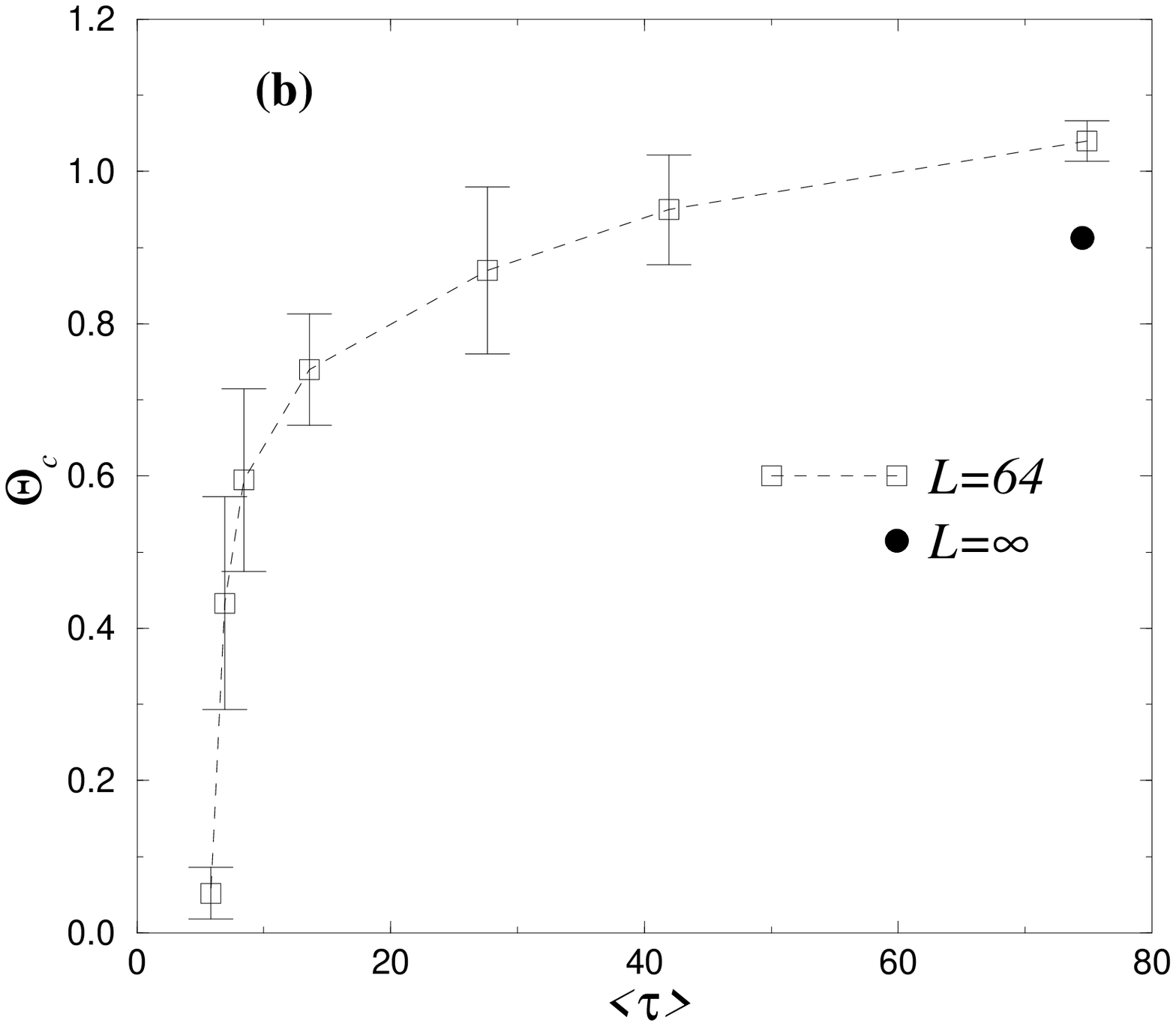}
\caption[]{
{\bf (a)} 
Order parameter $\langle |Q| \rangle$ vs. $\Theta$ 
for $L\!=\!64$ square system at 
$T \!=\! 0.8T_{\rm c}$ for several values of the field amplitude $H_0$. The 
corresponding metastable lifetimes, $\langle\tau(H_0,T)\rangle$ given in MCSS, 
are shown in the legends. 
{\bf (b)} 
The approximate dimensionless critical half-period 
$\Theta_{\rm c}$ vs. the metastable lifetime $\langle \tau(H_0,T) \rangle$
in MCSS}
\label{fig:Thetac}
\end{figure}

In terms of the dimensionless half-period, 
\begin{equation}
\label{eq:Theta}
\Theta = t_{1/2} \left/ \langle \tau(H_0,T) \rangle \right. \;,
\end{equation}
the DPT should occur at a critical value $\Theta_{\mathrm c}$ of order unity. 
Although $\Theta$ can be changed by varying either $t_{1/2}$, $H_0$, or $T$, 
in a first approximation we expect $\Theta_{\mathrm c}$ to be 
independent of $H_0$ and $T$. This expectation is confirmed by 
simulations carried out at several $H_0$ and $T$ for different system sizes. 
In particular, Fig.~\ref{fig:Thetac}(a) shows the average norm of the 
order parameter $\langle |Q|\rangle$ vs.\ $\Theta$ for
various field amplitudes and the corresponding metastable lifetimes. For 
weaker fields (longer lifetimes) the transition is apparent at some
$\Theta_{\mathrm c} \sim 1$, while it clearly disappears 
(no dynamically ordered phase exists) for 
sufficiently strong fields (small lifetimes). 
Figure~\ref{fig:Thetac}(b) shows 
$\Theta_{\mathrm c}$ vs.\ $\langle \tau(H_0,T) \rangle$ 
where $\Theta_{\mathrm c}$ was determined 
approximately as the location of the peak in the fluctuations of $Q$ in a 
$L\!=\!64$ system. Note that $\Theta_{\mathrm c}$ for an infinite system
is slightly different from this estimate due to finite-size effects.
Also, these effects are expected to be smaller for strong fields (small 
lifetimes) where the spins become uncorrelated. 
The deviations from unity for very small $\langle \tau(H_0,T) \rangle$ and the
complete disappearance of the DPT are discussed 
further in Sect.~\ref{sec:KJMA}.
 
In many studies of the DPT the transition has been approached 
by changing $H_0$ or $T$ \cite{LO90,ACHA95,ACHA97C,ACHA98}. 
While the above discussion indicates that this is correct in principle, 
we show in Sect.~\ref{sec:KJMA} 
that $\langle \tau(H_0,T) \rangle$ depends strongly and nonlinearly 
on its arguments. We therefore prefer changing 
$t_{1/2}$ at constant $H_0$ and $T$ \cite{SIDE99,SIDE98}, as this in practice 
gives more precise control over the distance from the transition. 

\section{Decay of Metastable Phases} 
\label{sec:KJMA} 
In mean-field studies of the DPT $H_0$ must be larger 
than the temperature-dependent spinodal
field beyond which the metastable 
free-energy minimum disappears \cite{JUNG90}. 
For weaker fields the hysteresis loops remain asymmetric, even for the 
lowest frequencies. It is much less appreciated that bounds 
on the fields and temperatures for which a DPT can occur also 
exist for spatially extended systems. These bounds are readily 
obtained from classical nucleation theory \cite{GUNT83B} and the 
Kolmogorov-Johnson-Mehl-Avrami (KJMA) theory of metastable decay 
\cite{SIDE99,SIDE98,RIKV94A,RIKV94,AVRAMI,RAMO99} by comparing 
four characteristic lengths. 
These are the lattice constant (here defined as unity), 
the size $R_{\rm c}(|H|,T)$ of 
a randomly nucleated critical droplet of stable phase,
the typical distance $R_0(|H|,T)$ between individual droplets, 
and the system size $L$. 

The critical radius is determined by the competition between the 
surface free energy of the droplet, $\propto \sigma(T) R^{d-1}$ 
where $\sigma(T)$ is the surface tension between the two phases, 
and the bulk free energy, $\propto - |H| R^d$. 
As a result, $R_{\rm c} \propto \sigma(T)/|H|$. The 
nucleation rate per unit volume is 
$I(H,T) \propto \exp \left[ - \Xi(T) / |H|^{d-1} \right]$, 
where $\Xi(T)$ is the field-independent part of the free energy of a 
critical droplet divided by $k_{\rm B} T$, 
and preexponential powers of $|H|$ have been suppressed.

The classical KJMA theory describes the metastable decay as homogeneous
nucleation of droplets at random times and positions with nucleation rate 
$I$, followed by deterministic growth of these droplets with 
constant radial velocity $v \propto |H|$. Assuming that the droplets overlap 
freely when they meet, the time dependent magnetization is given by the 
well-known ``Avrami's Law'',  
 \begin{eqnarray}
  m(t) &\approx& m_{\rm sp} \left\{ 2 \exp \left [- \int_{0}^{t}
          I \Omega_{d}(vt')^{d} {\rm d} t' \right ] - 1 \right\} \nonumber\\
       &=& m_{\rm sp} \left\{ 2  \exp \left [- \frac{\Omega_{d} v^{d} I}
                    {d+1} t^{d+1} \right ] -  1 \right\} \;,
 \label{eq:avrami}
 \end{eqnarray}
where $\Omega_{d}$ is a proportionality constant such that the volume 
of a droplet of radius $R$ equals $\Omega_{d} R^{d}$. 
The argument of the exponential is the ``extended volume'' \cite{AVRAMI}, 
i.e., the total volume fraction of stable-phase droplets, 
{\it uncorrected\/} for overlaps. Solving (\ref{eq:avrami}) 
for the time at which $m \! = \! 0$ gives the lifetime, 
\begin{equation}
\langle \tau (|H|,T) \rangle =
    \left [ \frac{\Omega_{d} v^{d} I}
            {(d+1) \ln 2 } \right ]^{-{1}/({d+1})} ,
\label{eq:tau_md}
\end{equation}
which depends on $|H|$ and $T$ through $v$ and $I$, but is 
{\it independent\/} of $L$. 
The characteristic length $R_0$ is obtained from 
$v(|H|,T)$ and $\langle \tau (|H|,T) \rangle$ as 
\begin{equation}
R_0(|H|,T) = 
v(|H|,T) \ \langle \tau (|H|,T) \rangle 
\propto 
\exp \left[\frac{\Xi(T)}{(d+1) |H|^{d+1}} \right] \;,
\label{eq:R0}
\end{equation}
where a preexponential power of $|H|$ again has been suppressed 
\cite{RIKV94A,RIKV94}. 

The regime in which a DPT can occur is that in which a large number of 
droplets contribute to the decay of the metastable phase: 
\begin{equation}
1 \ll R_{\rm c} \ll R_0 \ll L \;.
\label{eq:lengths}
\end{equation}
This is known as the multidroplet (MD) regime \cite{RIKV94A,RIKV94}. 
It is limited on the weak-field/small-system side by the 
{\it dynamic spinodal\/} (DSP) field , 
\begin{equation}
H_{\rm DSP}(T,L) \sim
\left( \frac{1}{d+1} \frac{\Xi(T)}{\ln L}\right)^{{1}/({d-1})} \;,
\label{eq:HDSP}
\end{equation}
which corresponds to $R_0 \! \approx \! L$. 
For $|H| < H_{\rm DSP}$ almost always 
only a {\it single\/} droplet contributes to the magnetization switching. 
A DPT does {\it not\/} occur in this regime, but 
{\it stochastic resonance\/}
is observed at low frequencies \cite{SIDE98A}. 
On the strong-field side the MD regime is limited by an $L$-independent 
cross-over field approximately given by $2 R_{\rm c} \! \approx \! 1$. 
By a somewhat confusing term this crossover is often referred to as  
the ``mean-field spinodal'' $H_{\rm MFSP}(T)$ \cite{RIKV94A,RIKV94}. 
In the {\it strong-field\/} (SF) regime beyond this limit 
the spins become increasingly uncorrelated as $|H|$ increases. 
This is the region of small $\langle \tau(H_0,T) \rangle$ 
(large $H_0$) where $\Theta_{\rm c}$ rapidly approaches zero and 
the transition disappears (Fig.~\ref{fig:Thetac}). 
In computational studies of the DPT it is essential to ensure that 
$H_{\rm DSP}(T,L) < H_0 < H_{\rm MFSP}(T)$ for all values of 
$L$ and $T$ used. 

\section{Results} 
\label{sec:res} 

We performed extensive simulations on square lattices 
with $L$ between 64 and 512 at $T \!=\! 0.8T_{\rm c}$ and $H_0 \!=\! 0.3J$. 
Typical run lengths near the transition range from 
$2$$\times$$10^4$ to $10^5$ full periods, corresponding to 
$2.8$$\times$$10^6-1.4$$\times$$10^7$ MCSS. The system was initialized
with all spins up and and the square-wave external field started with the
half-period in which $H\!=\!-H_0$. 
After some relaxation the system magnetization
reaches a limit cycle (except for thermal fluctuations), 
i.e., $Q$ is stationary. We discarded the first 500 periods of the 
time series to exclude transients from the stationary-state averages. 

\begin{figure}
\sidecaption
\includegraphics[width=.5\textwidth]{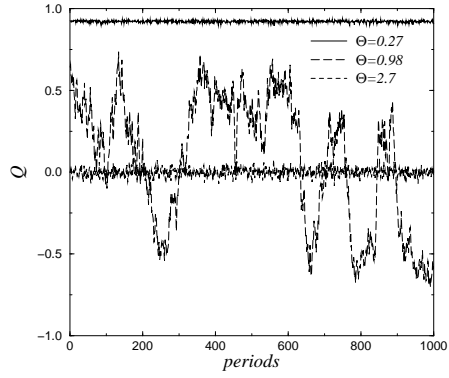} 
\caption[]{
Time series of the order parameter $Q$ 
at $T\!=\!0.8T_{\rm c}$ and $H_0\!=\!0.3J$ for $L\!=\!128$. 
Horizontal trace near $Q \!=\! +1$: 
$\Theta\!=\!0.27 < \Theta_{\rm c}$ 
(dynamically ordered phase). 
Strongly fluctuating trace: 
$\Theta\!=\!0.98 \! \approx \! \Theta_{\rm c}$ 
(near the DPT). 
Horizontal trace near $Q \!=\! 0$: 
$\Theta\!=\!2.7 > \Theta_{\rm c}$ 
(dynamically disordered phase)}
\label{fig:Qseries}
\end{figure}

\begin{figure}
\vspace{-1cm}
\hspace*{-1.5cm}
\includegraphics[width=.6\textwidth]{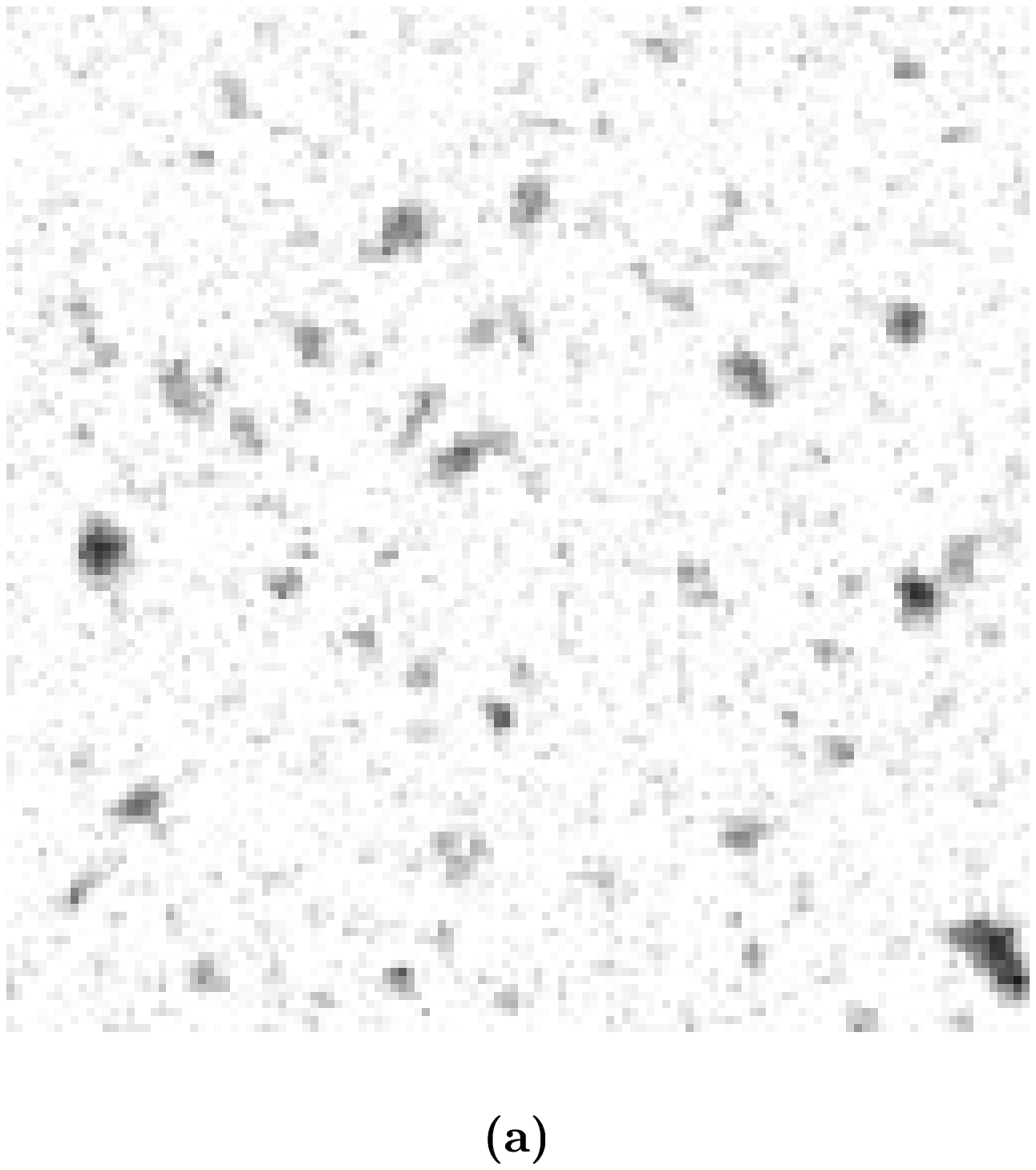} 
\includegraphics[width=.6\textwidth]{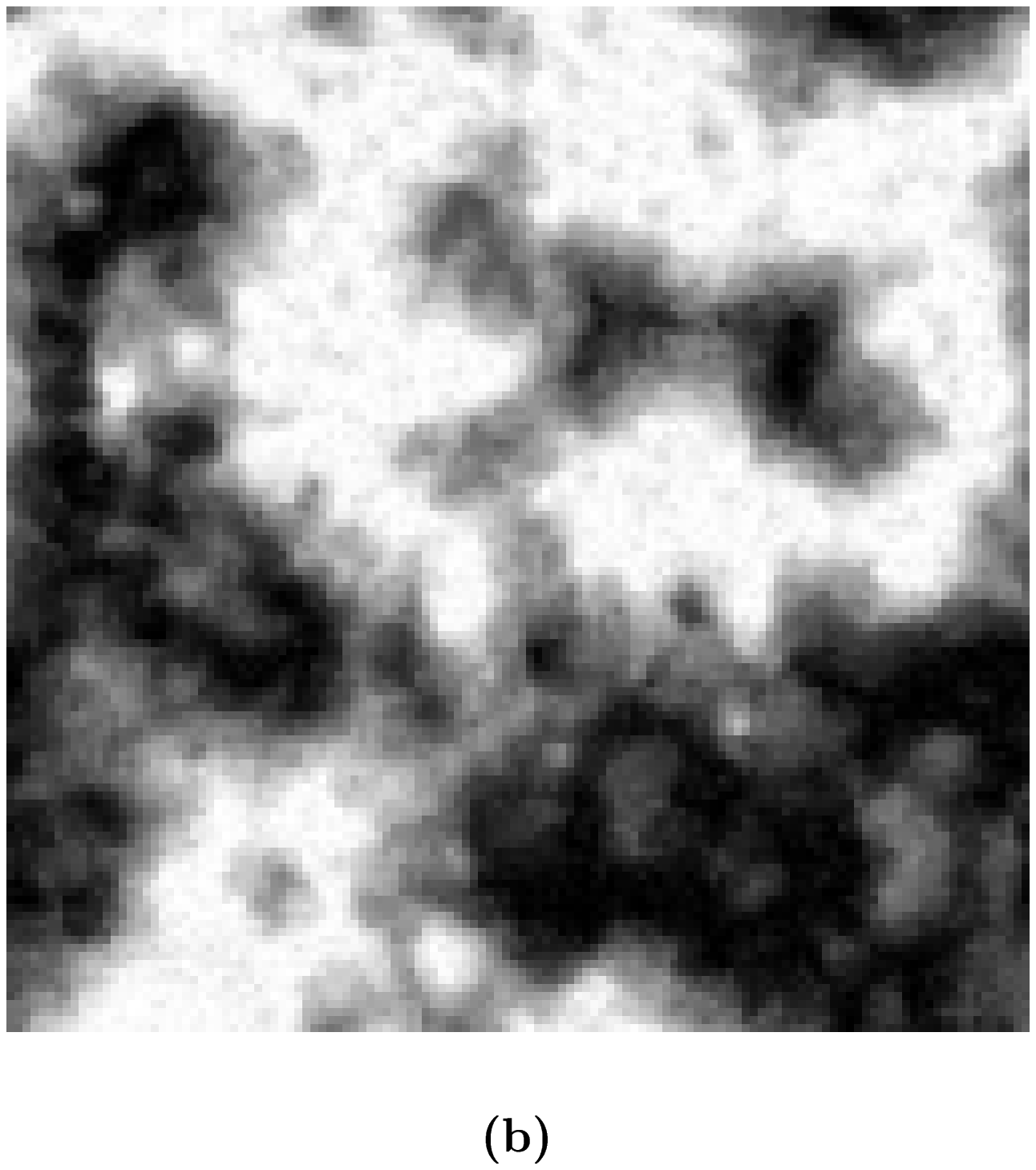} \\

\vspace{-4cm}
\hspace*{-1.5cm}
\includegraphics[width=.6\textwidth]{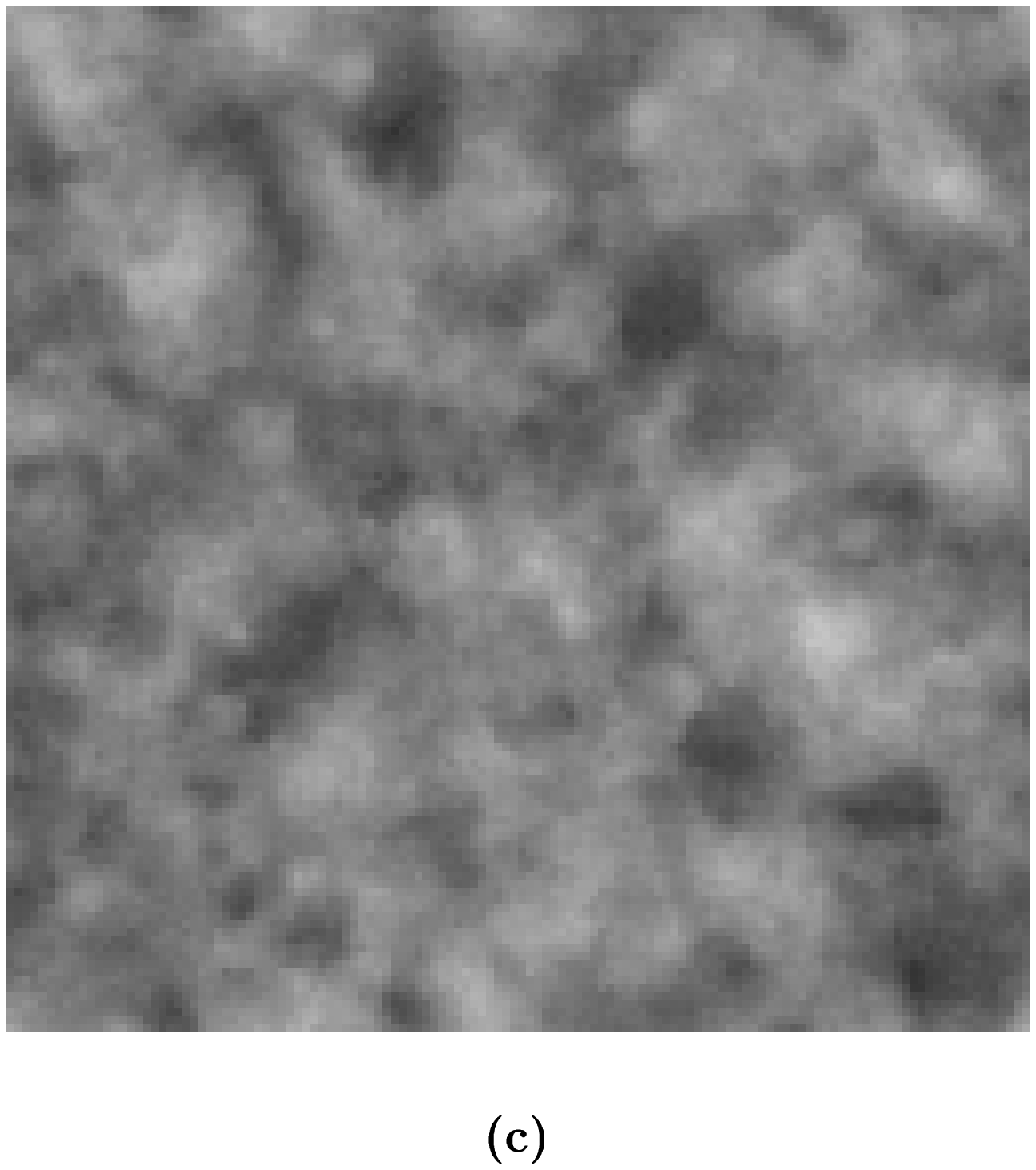}
\includegraphics[width=.6\textwidth]{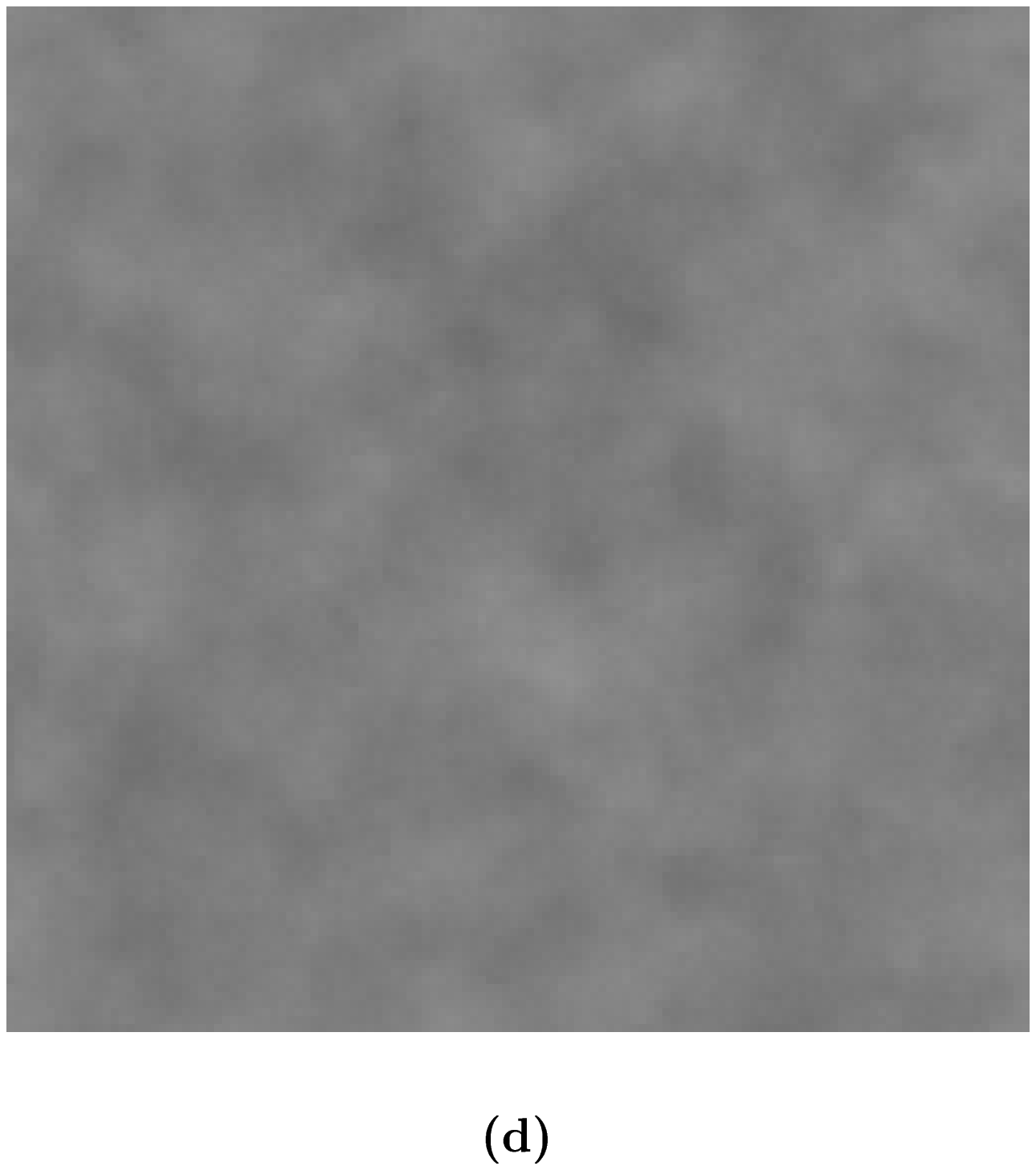} \\

\vspace{-10cm}
\begin{center} 
\includegraphics[width=.6\textwidth]{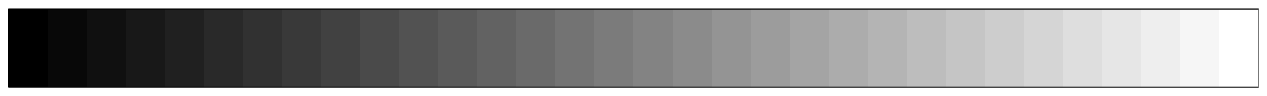}
\end{center} 

\vspace*{1cm}
\caption[]{ 
Configurations of the local order parameter $Q_i$ 
at $T\!=\!0.8T_{\rm c}$ and $H_0\!=\!0.3J$ for $L\!=\!128$. 
{\bf (a)} $\Theta\!=\!0.27 < \Theta_{\rm c}$ (dynamically ordered phase). 
{\bf (b)} $\Theta\!=\!0.98 \! \approx \! \Theta_{\rm c}$ 
(near the DPT).
{\bf (c)} $\Theta\!=\!2.7 > \Theta_{\rm c}$ 
(dynamically disordered phase).
{\bf (d)} $\Theta\!=\!13.4 \! \gg \! \Theta_{\rm c}$ 
(deeper in the disordered phase).
On the gray-scale $-1$ $(+1)$ corresponds to black (white)}
\label{fig:configs}
\end{figure}

For small half-periods ($\Theta\ll\Theta_{\rm c}$) 
the magnetization does not have time to 
switch, resulting in $|Q| \! \approx \! m_{\rm sp}$, 
while for large half-periods 
($\Theta\gg\Theta_{\rm c}$) it switches every half-period and 
$Q\! \approx \! 0$ 
as can be seen from the time series 
in Fig.~\ref{fig:Qseries}. The transition between the high- and low-frequency
regimes is singular, characterized by large fluctuations in $Q$.
To illustrate the spatial aspects of the transition 
we also show configurations of the local order parameter $Q_i$ 
in Fig.~\ref{fig:configs}. Below $\Theta_{\rm c}$ 
[Fig.~\ref{fig:configs}(a)] the majority of spins spend most
of their time in the $+1$ state, i.e., in the metastable phase during the 
first half-period, and in the stable 
equilibrium phase during the second half-period 
(except for equilibrium fluctuations). 
Thus, most of the $Q_i \! \approx \! +1$. 
Droplets of $s_i \!=\! -1$ that nucleate during the negative half-period and 
then decay back to $+1$ during the positive half-period show up as 
roughly circular gray spots in the figure. Since the spins near the 
center of such a droplet become negative first and revert to positive last, 
these spots appear darkest in the middle. 
Above $\Theta_{\rm c}$ 
[Figs.~\ref{fig:configs}(c,d)] the system follows the field
in every half-period (with some phase lag) and $Q_i \! \approx \! 0$ 
at all sites $i$. Near $\Theta_{\rm c}$ 
[Fig.~\ref{fig:configs}(b)] there are large clusters of
both $Q_i \! \approx \! +1$ and $Q_i \! \approx \! -1$ separated by wide 
``interfaces'' where $Q_i \! \approx \! 0$.
Also, for not too large lattices one often observes the full reversal of an 
ordered configuration $\{Q_i\}\rightarrow -\{Q_i\}$, typical of finite, 
spatially extended systems undergoing symmetry breaking.

\begin{figure}
\includegraphics[width=.48\textwidth]{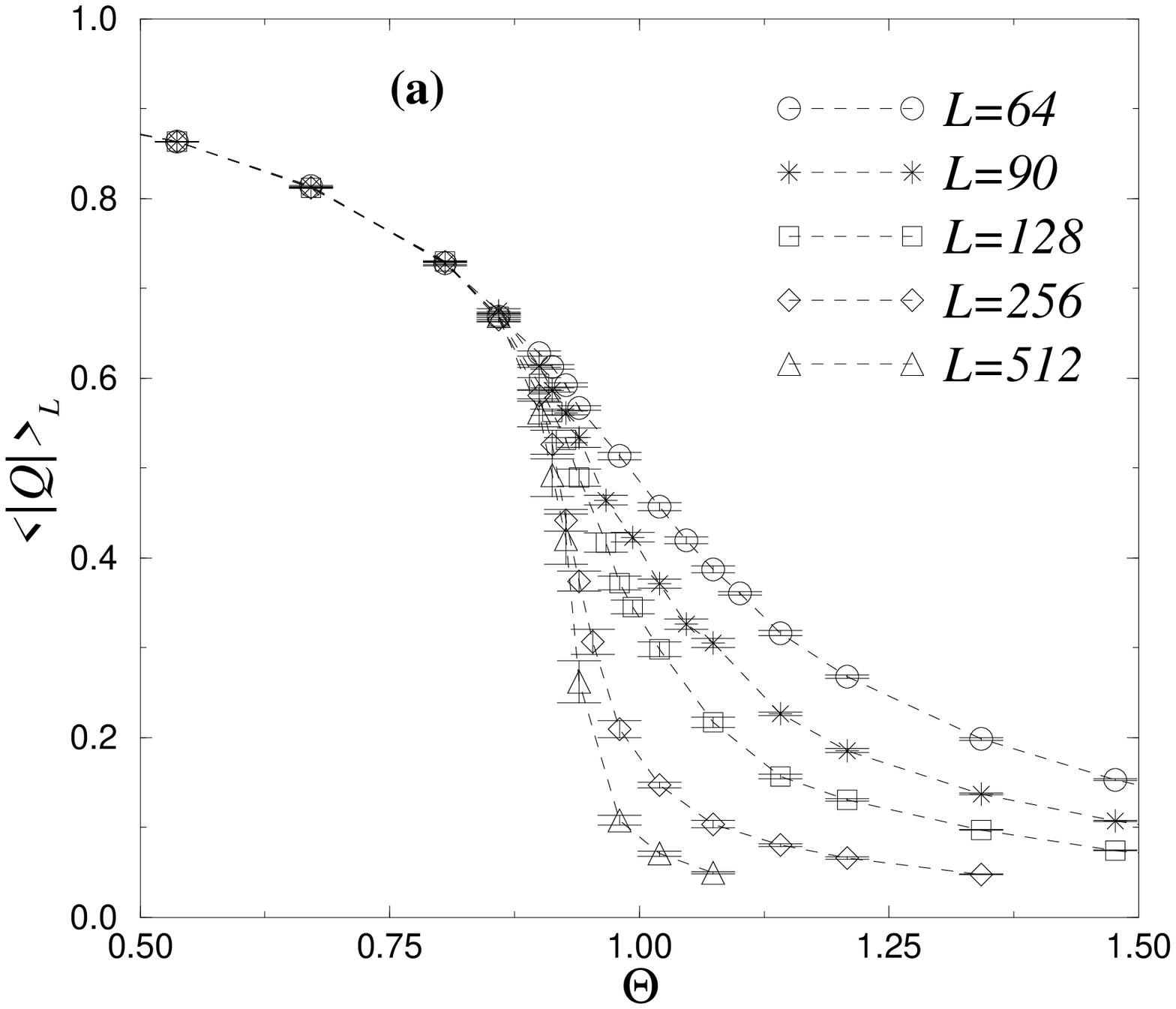} \hfill
\includegraphics[width=.48\textwidth]{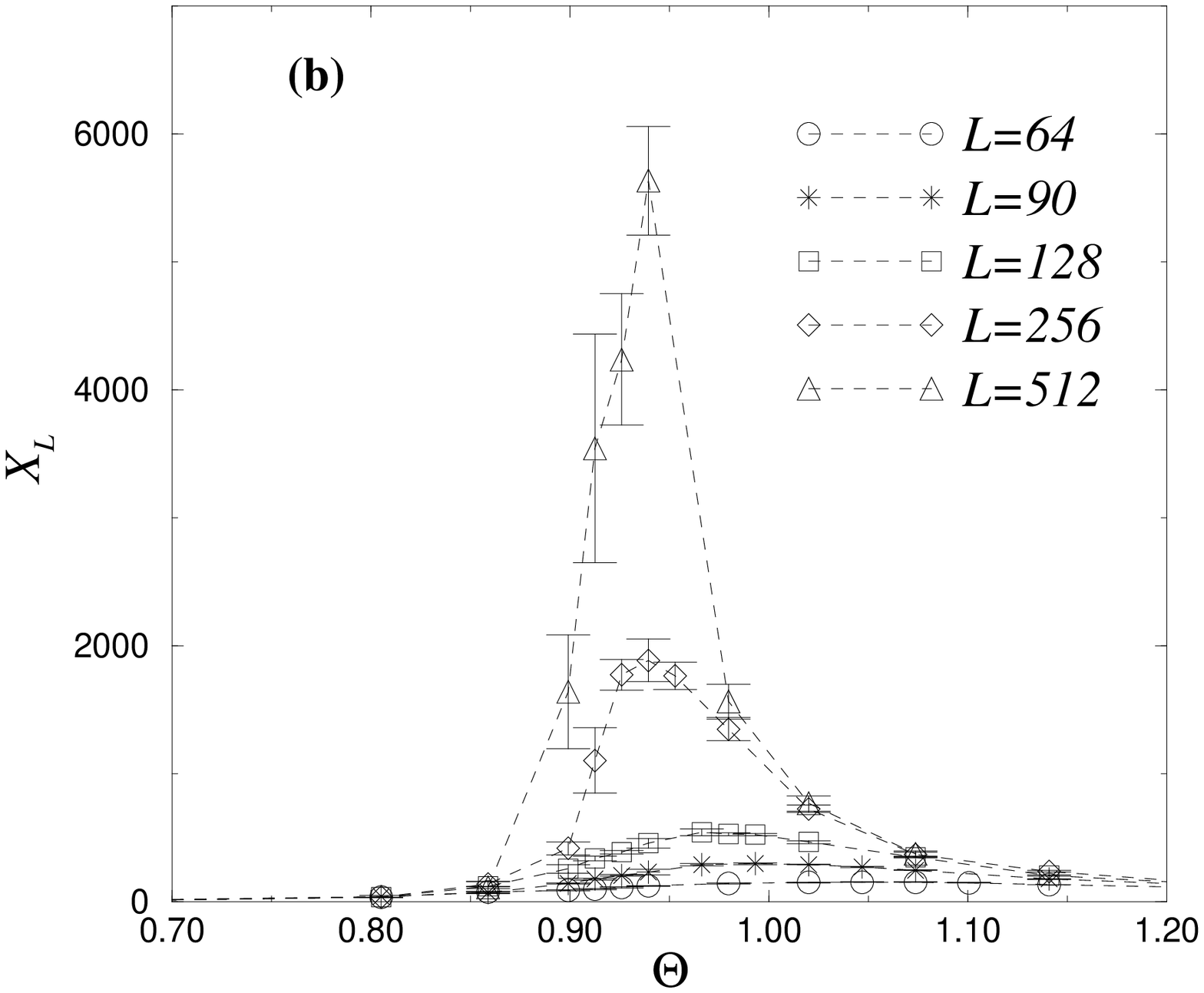} \hfill \\
\sidecaption
\includegraphics[width=.48\textwidth]{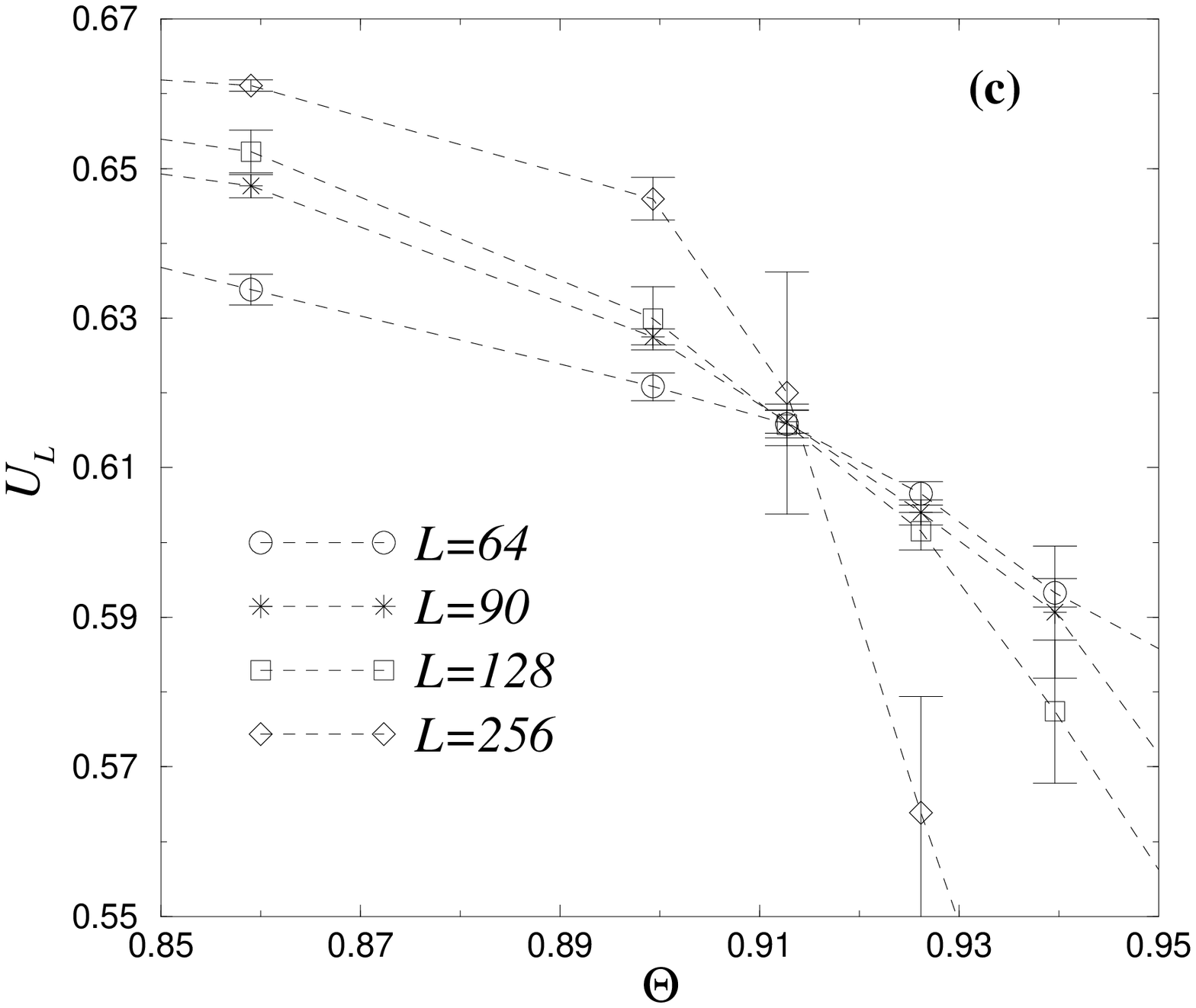}
\caption[]{
Finite-size behavior of our observables at $T\!=\!0.8T_{\rm c}$ and 
$H_0\!=\!0.3J$ for various system sizes.
{\bf (a)} The order parameter $\langle |Q|\rangle_L$.
{\bf (b)} The scaled variance $X_L$ as defined in (\protect\ref{eq:X}).
{\bf (c)} The fourth-order cumulant ratio as defined in 
(\protect\ref{eq:cumulant})}
\label{fig:fss}
\end{figure}

For finite systems in the dynamically 
ordered phase the distribution of $Q$ becomes 
bimodal. Thus, to capture symmetry breaking, one has to measure the average
norm of Q as the order parameter, i.e., $\langle |Q| \rangle$. 
Figure~\ref{fig:fss}(a) clearly shows that this order parameter is of order
unity for $\Theta<\Theta_{\rm c}$ and vanishes for $\Theta>\Theta_{\rm c}$, 
except for finite-size effects. 

To characterize and quantify this transition in terms of critical exponents
we employ the well-known technique of finite-size 
scaling \cite{BIND92B,BIND90}.
The quantity analogous to the susceptibility is the scaled variance
of the dynamic order parameter, 
\begin{equation}
X_L=L^2 \left(\langle Q^2 \rangle_L - \langle |Q| \rangle_L ^2 \right) \; .
\label{eq:X}
\end{equation}
For finite systems $X_L$ has a characteristic peak near $\Theta_{\rm c}$ 
[see Fig.~\ref{fig:fss}(b)] 
which increases in height with increasing $L$, while no finite-size effects
can be observed for $\Theta\ll\Theta_{\rm c}$ 
and $\Theta\gg\Theta_{\rm c}$. This implies
the existence of a divergent length scale, possibly the 
correlation length which governs the long-distance behavior of the local
order-parameter correlations $\langle Q_i Q_j \rangle$. Note that the location
of the maximum in $X_L$ shifts with $L$. 
This also contains important information about the critical exponents.

To estimate the value of 
$\Theta_{\rm c}$ at which the transition occurs in an infinite system
we use the intersection of the fourth-order 
cumulant ratios \cite{BIND92B,BIND90}, 
\begin{equation}
U_L=1 - \frac{\langle Q^4\rangle_L}{3\langle Q^2\rangle_L^2} \;,
\label{eq:cumulant}
\end{equation}
for several system sizes as shown in Fig.~\ref{fig:fss}(c). 
For the largest system ($L \!=\! 512$) 
the error bars on $U_L$ were too large to 
use it to obtain estimates for the crossing. Our estimate for the 
dimensionless critical half-period is 
$\Theta_{\rm c} \!=\! 0.913 \pm 0.003$ with a fixed-point value 
$U^* \!=\! 0.615 \pm 0.005$ for the cumulant ratio. 

For our model the quantity analogous to the reduced temperature in 
equilibrium systems (i.e., the distance from the critical point) is
\begin{equation}
\theta= \frac{|\Theta-\Theta_{\rm c}|}{\Theta_{\rm c}}\;.
\label{eq:reduced_Theta}
\end{equation}
Finite-size scaling theory provides simple relations for the
order parameter and its scaled variance $X_L$ for finite systems in the 
critical regime \cite{BIND92B,BIND90}:
\begin{subeqnarray}
\langle |Q|\rangle _L 
& = & 
L^{-\beta /\nu} {\cal F}_{\pm}(\theta L^{1/\nu}) \;, 
\label{eq:Qscaling1} \\ 
X_L 
& = & 
L^{\gamma /\nu} {\cal G}_{\pm}(\theta L^{1/\nu})\;, 
\label{eq:Xscaling1}
\end{subeqnarray}
where ${\cal F}_{\pm}$ and ${\cal G}_{\pm}$ are 
scaling functions and the $+$ ($-$) 
index refers to \mbox{$\Theta >\Theta_{\rm c}$} 
\mbox{($\Theta <\Theta_{\rm c}$)}.
Then at $\Theta_{\rm c}$ ($\theta\!=\!0$) 
we have the finite-size behavior of the above two observables:
\begin{subeqnarray}
\langle |Q|\rangle _L & \propto & L^{-\beta /\nu} \;, \label{eq:Qscaling2} \\
X_L & \propto & L^{\gamma /\nu} \;.
\label{eq:Xscaling2}
\end{subeqnarray}
Using $\langle |Q|\rangle _L$ and  $X_L$ for several system sizes at
$\Theta_{\rm c}$ 
(estimated as the value of $\Theta$ where the cumulants cross) 
we employ (\ref{eq:Qscaling2}) 
and (\ref{eq:Xscaling2}) to find the exponent ratios $\beta /\nu$ and 
$\gamma /\nu$ through weighted linear least-squares fitting to the 
logarithmic data. 
Further, the shift in the location of the peak in $X_L$ for 
these finite systems, $\Theta_{\rm c}(L)$, is used to estimate $\nu$ 
in the same way \cite{BIND92B,BIND90}:
\begin{equation}
|\Theta_{\rm c}(L)-\Theta_{\rm c}| \propto L^{-1/\nu} \;.
\label{eq:Theta_cscaling}
\end{equation}
Our estimates for these exponents are $\beta/\nu$$=$$0.122\pm 0.005$, 
$\gamma/\nu$$=$$1.77\pm 0.05$, and $\nu$$=$$1.0\pm 0.15$. 
The largest uncertainty occurs in 
$\nu$ due to the large relative errors in 
$|\Theta_{\rm c}(L)-\Theta_{\rm c}|$. 
We note that these numbers are very close (within the one-standard-deviation 
error bars) 
to the critical exponents of the two-dimensional Ising universality class 
$\beta/\nu$$=$$1/8$$=$$0.125$, $\gamma/\nu$$=$$7/4$$=$$1.75$, and 
$\nu$$=$$1$. 
However, as discussed in Sect.~\ref{sec:disc}, we do not consider this 
conclusive proof of the universality class for this nonequilibrium 
phase transition. 

\begin{figure}
\includegraphics[width=.45\textwidth]{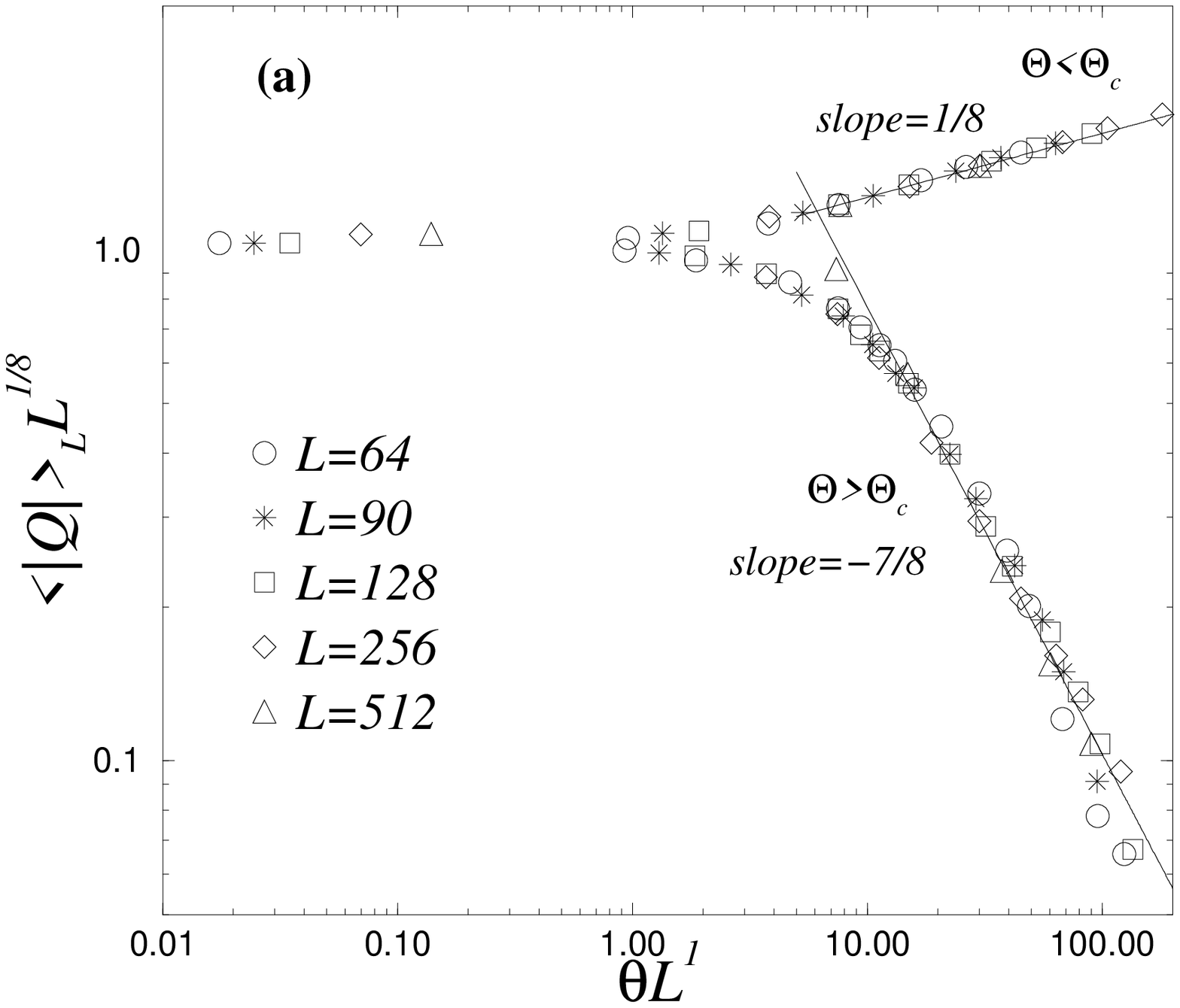} \hfill
\includegraphics[width=.45\textwidth]{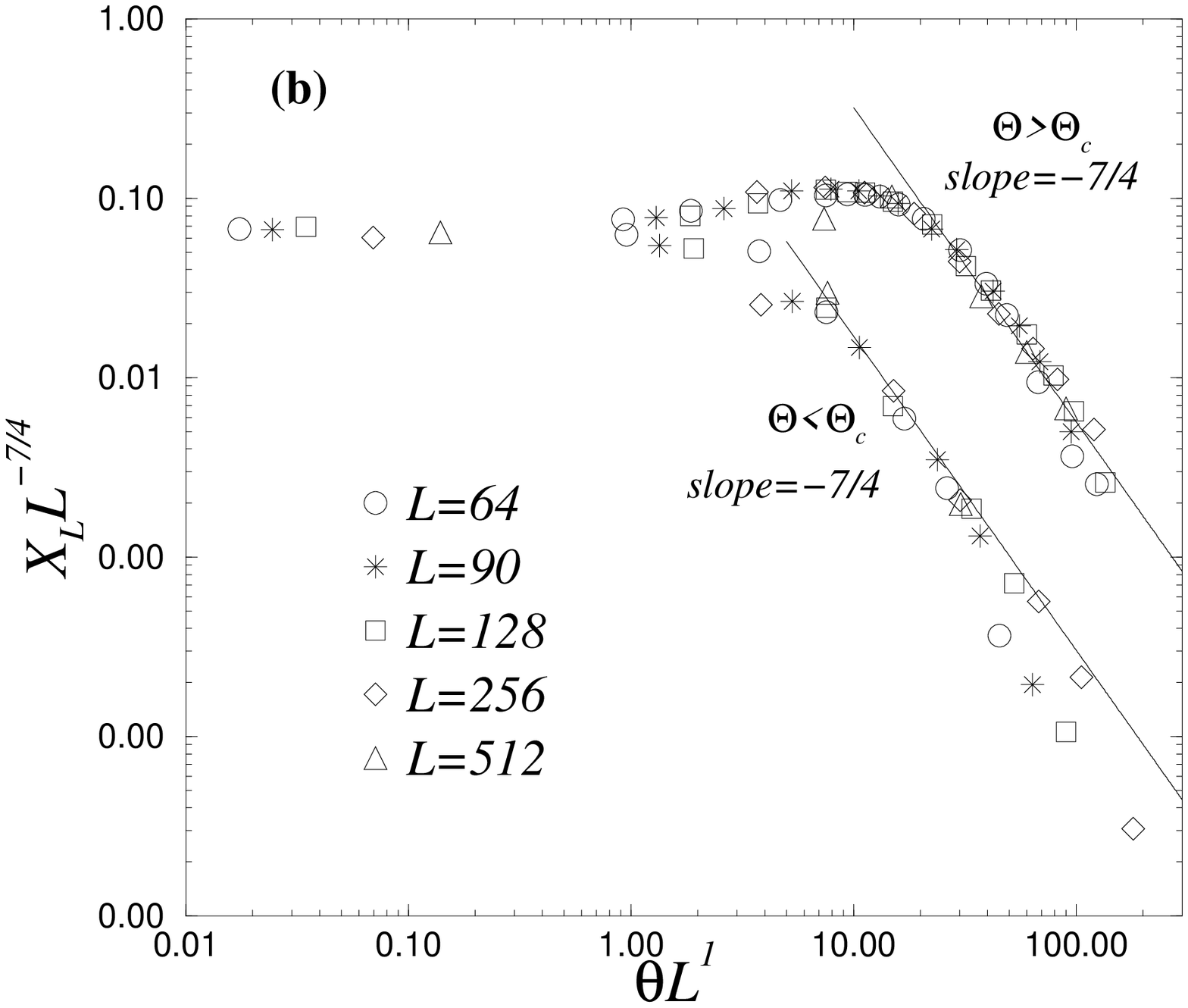} \\
\includegraphics[width=.45\textwidth]{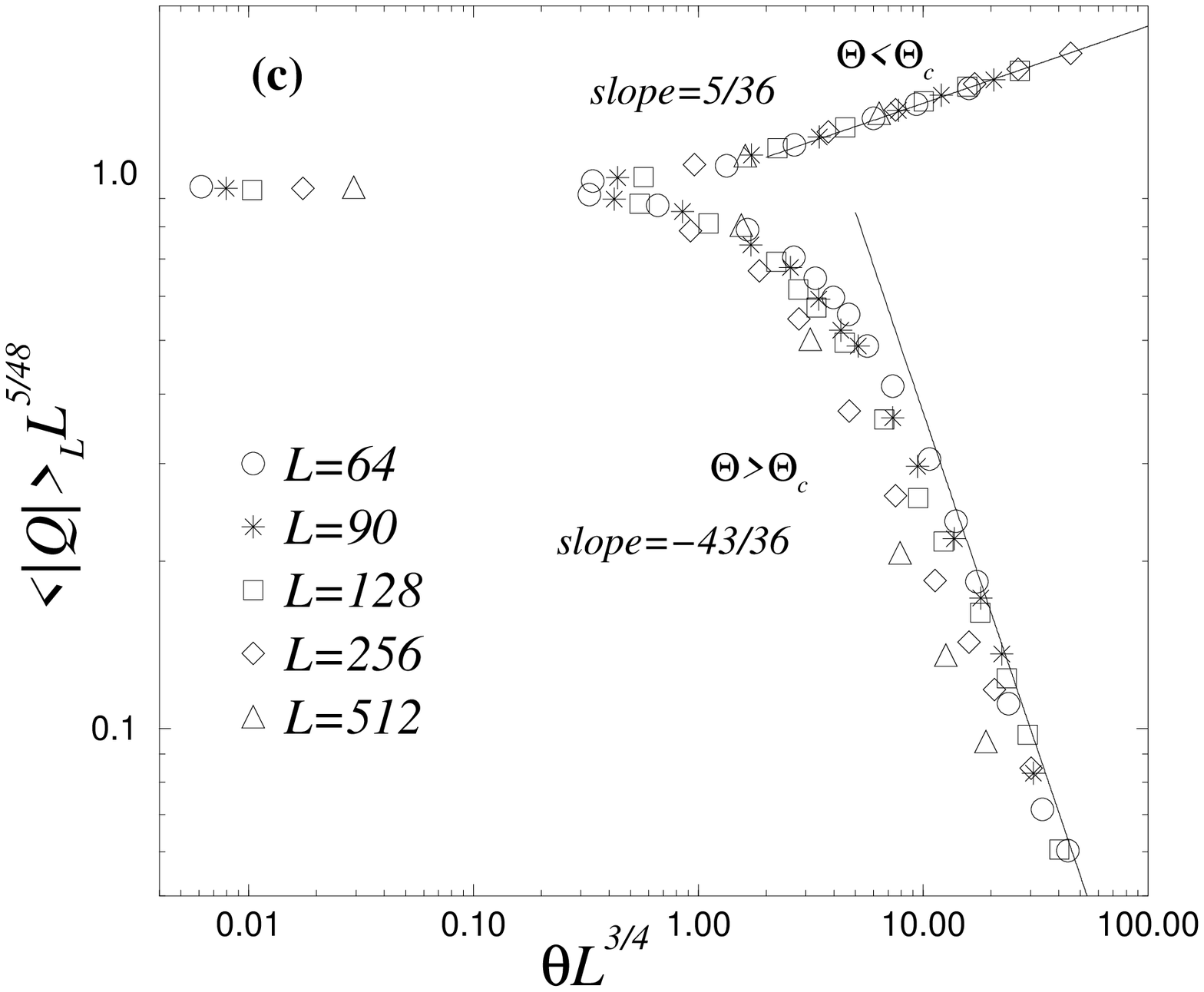} \hfill
\includegraphics[width=.45\textwidth]{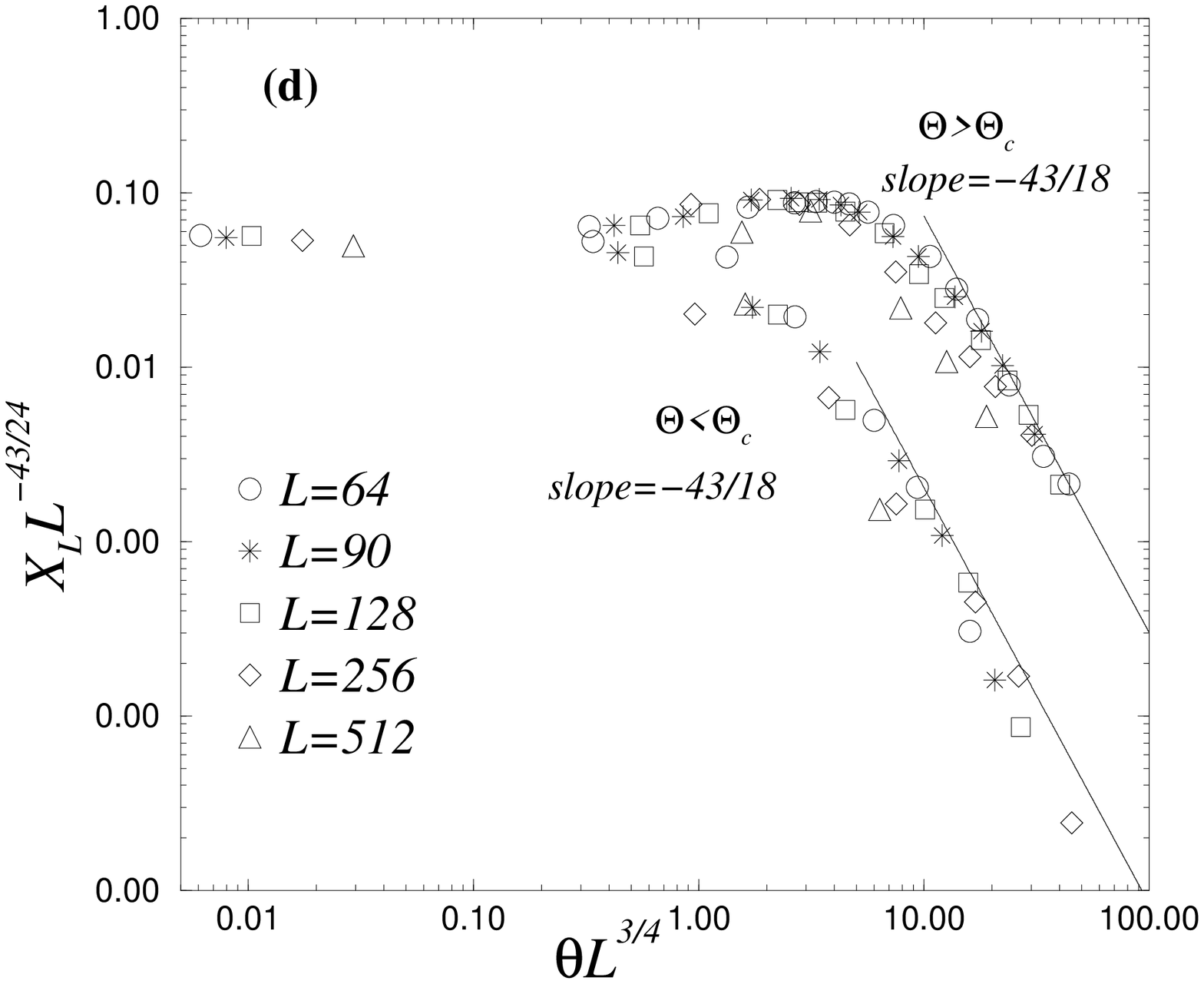}
\caption[]{
Finite-size scaling plots at $T\!=\!0.8T_{\rm c}$ and $H_0\!=\!0.3J$ using the 
two-dimensional Ising exponents: 
{\bf (a)} For the order parameter $\langle |Q| \rangle_L$. 
{\bf (b)} For the scaled variance $X_L$.
Finite-size scaling plots at the same temperature and field using the
random-percolation exponents:
{\bf (c)} For the order parameter $\langle |Q| \rangle_L$. 
{\bf (d)} For the scaled variance $X_L$.
The straight lines in all graphs represent the asymptotic 
large-argument behaviors of the functions
${\cal F}_{\pm}$ and ${\cal G}_{\pm}$ given in 
(\protect\ref{eq:Qscaling1}) and (\protect\ref{eq:Xscaling1}), 
respectively. The value for the (infinite-system) $\Theta_c$, used in all four
graphs, was obtained from the cumulant crossing 
[(\protect\ref{eq:cumulant})] Fig.~\ref{fig:fss}(c)}
\label{fig:ising_collapse}
\end{figure}

To obtain a more general picture of how well the scaling 
relations in (\ref{eq:Qscaling1}) and 
(\ref{eq:Xscaling1}) are obeyed, we plot $\langle |Q|\rangle _L L^{\beta/\nu}$
[Fig.~\ref{fig:ising_collapse}(a)] and $X_L L^{-\gamma/\nu}$ 
[Fig.~\ref{fig:ising_collapse}(b)] vs. $\theta L^{1/\nu}$ \cite{LANDAU76}. 
In these figures we used the exponents of the two-dimensional 
Ising universality class since they are within one standard deviation 
of our estimates and analogous data-collapse plots using our numerical 
exponent estimates do not look perceptibly different. The plots show excellent
agreement with the scaling assumption and graphically define the 
scaling functions
${\cal F}_{\pm}$ and ${\cal G}_{\pm}$. The asymptotic behaviors of these 
functions have to be simple power laws so that the true critical behavior is
restored when the limit $L\rightarrow\infty$ is taken in 
(\ref{eq:Qscaling1}) and (\ref{eq:Xscaling1}).
Some deviation can be observed in Fig.~\ref{fig:ising_collapse}(b): 
below $\Theta_{\rm c}$ 
the data points for the smaller systems systematically start 
to peel off earlier from the straight line representing the asymptotic 
behavior of ${\cal G}_{-}$ for large argument. 

In the absence of theoretical arguments why this DPT should belong to the
Ising universality class we also tested the data collapse with the 
exponents for random percolation, which are relatively close to the 
corresponding Ising ones
\cite{STAU92}: $\beta/\nu$$=$$5/48$$\approx$$0.104$, 
$\gamma/\nu$$=$$43/24$$\approx$$1.79$, and $\nu$$=$$4/3$$\approx$$1.33$.
It appears reasonable to consider this universality class in particular, 
since the metastable decay process described in Sect.~\ref{sec:KJMA} 
produces transient spanning clusters that belong to the random-percolation 
class as $m(t)$ passes through $H$- and $T$-dependent percolation thresholds 
for the two phases near $m \! = \! 0$ \cite{SIDEUNP}. 
Data-collapse plots using the random-percolation exponents are
shown in Figs.~\ref{fig:ising_collapse}(c,d). They are clearly inferior 
to the ones with the Ising exponents [Figs.~\ref{fig:ising_collapse}(a,b)].
In particular, for $\Theta >\Theta_c$ (the dynamically disordered phase) 
the data collapse gets progressively worse for larger systems. 
Comparing the scaling plots in Figs.~\ref{fig:ising_collapse}(a,b) with 
Figs.~\ref{fig:ising_collapse}(c,d) we conclude that the exponent $\nu$
for the DPT is significantly different from that of the random percolation 
universality class and that it is closer to the Ising value. This 
conclusion is 
supported by direct comparison of the random-percolation and Ising 
exponents with our numerical estimates for the DPT. Our estimates for 
$\beta/\nu$ and $\nu$ lie more than two standard deviations away from the 
percolation values, but less than one standard deviation away from the 
Ising values, while our estimate for $\gamma/\nu$ lies midway between and 
less than one standard deviation away from the exact values for both classes. 

The exponents previously obtained with a sinusoidally 
oscillating field \cite{SIDE99,SIDE98A} lie within two 
standard deviations of our values.  
As it was difficult to determine the uncertainties in 
the estimates obtained in that 
study, we believe the results are consistent and indicate that the 
DPT in systems driven by sinusoidal and square-wave 
fields belong to the same universality class, as expected.

\section{Summary and Discussion} 
\label{sec:disc} 
In this paper we have presented preliminary results from a 
large-scale Monte Carlo study of the dynamic phase transition (DPT) in 
a two-dimensional kinetic Ising ferromagnet driven by a square-wave 
oscillating field. Our results are consistent with 
those of previous studies of 
the same model in a sinusoidally varying field \cite{SIDE99,SIDE98}. 
They indicate that for 
field amplitudes $H_0$ such that the metastable magnetized phase decays 
to equilibrium via the multidroplet (MD) mechanism described in 
Sect.~\ref{sec:KJMA}, the system undergoes a continuous DPT 
when the half-period of the field, $t_{1/2}$, 
approximately equals the metastable lifetime, $\langle \tau(H_0,T) \rangle$. 
Thus the critical value $\Theta_{\rm c}$ of the dimensionless  
half-period defined in (\ref{eq:Theta}) is near unity. 
As $\Theta$ is increased past $\Theta_{\rm c}$ the order parameter 
$\langle |Q| \rangle$, 
which is the expectation value of the norm of the 
period-averaged magnetization, vanishes in a singular fashion, as 
shown in Fig.~\ref{fig:fss}(a). 

The strong and systematic finite-size 
effects in the order parameter and its scaled variance indicate that 
there is a divergent correlation length associated with the transition. 
Using standard finite-size scaling techniques borrowed from the theory 
of equilibrium phase transitions we therefore estimated $\Theta_{\rm c}$ 
and the critical exponents $\beta$, $\gamma$, and $\nu$ from data for 
system sizes between $L \!=\! 64$ and 512. The resulting estimates are 
$\beta/\nu$$=$$0.122\pm 0.005$, $\gamma/\nu$$=$$1.77\pm 0.05$, 
and $\nu$$=$$1.0\pm 0.15$, which agree to within the statistical 
errors with those previously obtained 
with a sinusoidally oscillating field \cite{SIDE99,SIDE98A}. 
This is strong evidence 
that the shape of the wave does not effect the universal aspects of the DPT.

We also note that 
our estimates agree well within one standard deviation 
with the exact values for the 
two-dimensional equilibrium Ising model: 
$\beta/\nu$$=$$1/8$$=$$0.125$, $\gamma/\nu$$=$$7/4$$=$$ 1.75$, 
and $\nu$$=$$1$, 
and that they satisfy the exponent relation 
\begin{equation}
2 \left( \beta / \nu \right) + \left( \gamma / \nu \right) 
= 2.01 \pm 0.05 \approx d \;. 
\label{eq:hyper}
\end{equation}
The fixed-point value of the fourth-order cumulant ratio, 
$U^* \!=\! 0.615 \pm 0.005$, is also close to that of the Ising model, 
$U^* \!=\! 0.610 \ 690 \ 1(5)$ \cite{KAMI93}. 
As discussed in \cite{SIDE99} we find the 
estimated values for the exponents and $U^*$ conclusive evidence 
that the DPT corresponds to a nontrivial fixed point, so that 
(\ref{eq:hyper}) indeed represents a hyperscaling relation. 

Since we know of no convincing theoretical argument that this 
nonequilibrium phase transition should be in any particular equilibrium 
universality class, we also compare the exponent estimates with 
those of the universality class 
of random percolation. For this class the values \cite{STAU92}
$\beta/\nu$$=$$5/48$$\approx$$0.104$ and $\nu$$=$$4/3$$\approx$$1.33$ are 
significantly different from our exponent estimates, while 
$\gamma/\nu$$=$$43/24$$\approx$$1.79$ lies within the uncertainty in our 
estimate. 
Taken together, the three estimates and the low quality of the data 
collapse plots shown in Figs.~\ref{fig:ising_collapse}(c,d) 
constitute strong evidence that 
the DPT is {\it not} in the universality class of random percolation. 
However, we emphasize that we do not yet consider the question of the 
universality class as settled. That will 
require still more accurate numerical exponent estimates 
as well as theoretical arguments. 

Outside the MD regime the phase transition disappears. For weaker fields 
or smaller systems the metastable phase decays stochastically through 
a single droplet, and no long-range correlations evolve \cite{SIDE98A}. 
{}In the other extreme, as $H_0$ is increased into the strong-field 
regime, $\Theta_{\rm c}$ goes to zero and vanishes at a sharply defined 
field (lifetime), as shown in Fig.~\ref{fig:Thetac}(b). 

In conclusion, the dynamic phase transition observed in spatially extended 
kinetic Ising systems driven by oscillating fields is a fascinating 
nonequilibrium critical phenomenon with all the hallmarks of a thermal 
phase transition corresponding to a nontrivial fixed point. 
Although the present study confirms that 
the particular wave form of the oscillating field does not change the 
universality class, and our numerical estimates for the critical exponents 
and cumulant ratio are consistent 
with the universality class of the two-dimensional Ising model in 
equilibrium, many questions remain. These include the question of 
universality under changes in $T$ and $H_0$, and the 
details of the disappearance of the transition as $H_0$ is 
increased. No procedure to identify critical correlated clusters from the 
connected clusters seen in configuration ``snapshots'' such as 
Fig.~\ref{fig:configs}(b) is yet known, 
and theoretical ideas are largely missing. 
Thus, there is still much to be done! 

\section*{Acknowledgments} 
\label{sec:ack} 
This research was supported in part by Florida State University through 
the Center for Materials Research and Technology and the Supercomputer
Computations Research Institute (partially funded by the
US Department of Energy through Contract No.\ DE-FC05-85ER25000), 
and by the National Science Foundation through Grants No.\ 
DMR-9634873 and DMR-9871455. Computing resources at the 
National Energy Research Scientific Computing Center 
were made available by the US Department of Energy under Contract No.\
DE-AC03-76SF00098.


\clearpage

\flushbottom

\end{document}